\documentclass[sigconf,nonacm]{acmart}
\AtBeginDocument{%
  }

\setcopyright{acmlicensed}
\copyrightyear{2026}
\acmYear{2026}
\acmConference[RecSys '26]{20th ACM Conference on Recommender Systems}{September 28--October 2, 2026}{Minneapolis, MN, USA}

\usepackage{subcaption}

\begin{document}

\title{Book Readership During Movie Releases: An Exploratory Analysis}

\author{Sushobhan Parajuli}
\email{sp3886@drexel.edu}
\orcid{0009-0008-5679-9524}
\affiliation{%
  \institution{Dept. of Information Science \\ Drexel University}
  \city{Philadelphia}
  \state{PA}
  \country{USA}
}

\author{Vittoria Vineis}
\email{vineis@diag.uniroma1.it}
\authornote{Work carried out while visiting Drexel University.}
\orcid{0009-0003-6074-8344}
\affiliation{%
  \institution{Sapienza University of Rome}
  \city{Rome}
  \country{Italy}
}

\author{Samira Vaez Barenji}
\email{svaez@drexel.edu}
\orcid{0000-0002-2123-4338}
\affiliation{%
  \institution{Dept. of Information Science \\ Drexel University}
  \city{Philadelphia}
  \state{PA}
  \country{USA}
}

\author{Michael D. Ekstrand}
\email{mdekstrand@drexel.edu}
\orcid{0000-0003-2467-0108}
\affiliation{%
  \institution{Dept. of Information Science \\ Drexel University}
  \city{Philadelphia}
  \state{PA}
  \country{USA}
}
\renewcommand{\shortauthors}{Parajuli et al.}

\begin{abstract}
Exogenous events can temporarily change the relevance of items in recommender systems, but these shifts are often not visible in historical interaction data until after users have already responded.
In book recommendation, movie adaptations provide a clear example of such events: the release of a movie based on a book can temporarily increase attention to the source text and change its relevance for some readers. 
We examine this phenomenon using a large Goodreads dataset matched to movie release dates.
We find a clear spike in readership around the release month, and then we evaluate existing recommendation models to understand how they rank movie-adapted books around the movie release date.

\end{abstract}




\maketitle

\section{Introduction}
Recommender systems have incorporated item-related information beyond user-item interactions to improve recommendations, including item attributes, knowledge graphs, and temporal popularity signals \citep{sun2019research, koren2009collaborative}.
One type of information that has received little attention is exogenous events tied to specific items.
Movie adaptations of books are a natural example: when a movie based on a book is released, the original book can receive renewed attention from new readers, returning fans, and audiences comparing the adaptation with the source material.

\begin{figure}[ht!]
    \centering
\includegraphics[width=0.9\linewidth]{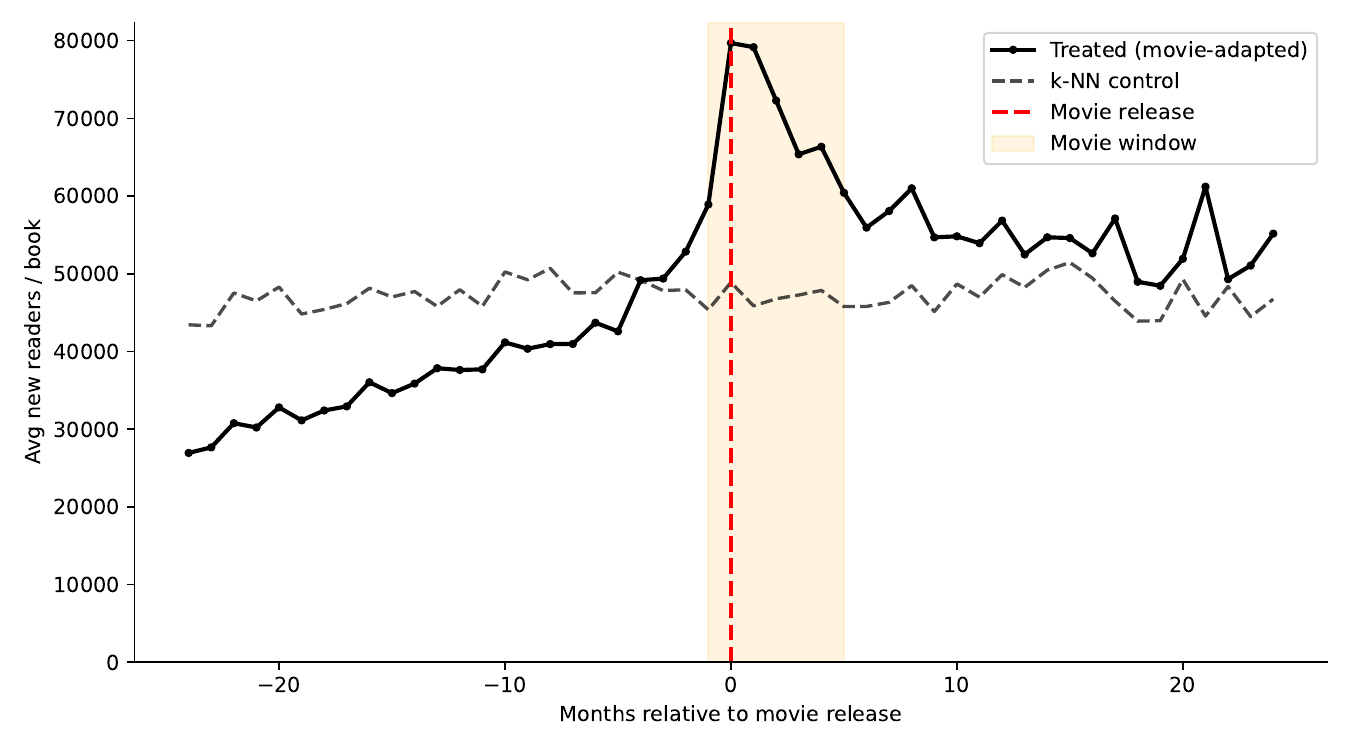}
\caption{Pooled event study across all movie-adapted books: average new readers
per book by month relative to film release ($t{=}0$, red dashed), for treated
books (solid) and their k-NN matched controls (dashed).}
\label{fig:es-overall}
\end{figure}

Prior studies in economics and information science suggest that this shift in demand is real.
Promotional events have been shown to drive demand for related products --- endorsements lift sales of associated titles \citep{garthwaite2014demand}, and new releases increase interest in back-catalog items \citep{hendricks2009information}.
Closer to our setting, studies of book sales and library circulation have found increases in book interest around movie releases, with effects strongest immediately before and after release \citep{Montesi2014FilmAdaptation, ponzoHaveYouRead}, as also illustrated in Figure \ref{fig:es-overall}.
These findings suggest that movie releases result in a meaningful, time-bounded signal for book demand that occurs before sufficient new interaction data has accumulated for standard recommender models to respond.

Yet to our knowledge, no prior work has analyzed whether existing recommender models naturally surface movie-adapted books during their release windows.
Before attempting to incorporate such signals into a model, it is worth asking whether standard collaborative filtering models already capture this effect implicitly through interaction patterns.
In this paper, we study movie adaptations as exogenous events in book recommendation.
Using movie-book pairs from Wikidata, movie metadata from TMDB, and Goodreads interaction data, we address two research questions:

\begin{description}
\item[RQ1] To what extent are movie adaptations associated with changes in book readership, and how does this association vary across books with different popularity levels?
\item[RQ2] Do recommender models assign systematically different scores to books with a recent or imminent movie adaptation, and how does this vary across recommendation models?
\end{description}

\begin{figure*}[ht]
    \centering
\includegraphics[width=\linewidth]{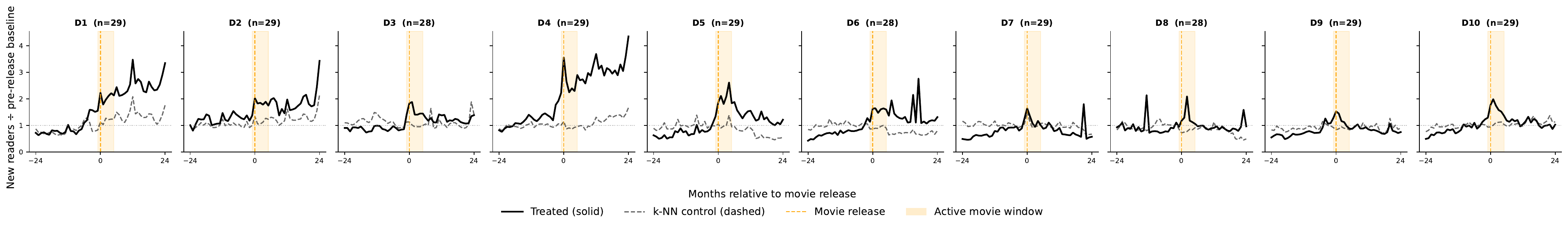}
\caption{Average monthly new Goodreads readers per movie-adapted book (solid line) and k-NN matched controls (dashed line), plotted relative to the film release date (month 0), stratified by pre-release book readership deciles (ascending order). Within each decile, both curves are normalized by the control group's mean
readership over the immediate pre-release window.}
\label{fig:event_study}
\end{figure*}

\section{Data}
We use the UCSD Book Graph \cite{wan2018item, wan2019fine}, augmented by \citet{ekstrand2018exploring}, as the source of user-book interactions.
The dataset contains over 228 million user-book interactions across 876,145 users and 1.52 million books, spanning January 2007 to November 2017.


We construct a dataset of movie-book pairs by querying Wikidata for US movies (P31: Q11424, P495: Q30) with a ``based on'' relationship to a book (P144), released between 2007 and 2017.
We filter to movies with at least 1,000 cumulative TMDB votes as a proxy for cultural impact. 
We link adapted books to Goodreads records via Wikidata work ID (P8383), fuzzy title matching, and manual verification for a small number of remaining cases.
This yields 287 movie-book pairs released between 2007-01-01 and 2017-12-31, each with movie and book identifiers, release date, and TMDB metadata.
We release this dataset along with the code.\footnote{The code is available at \url{https://zenodo.org/records/21464427}}

Finally, we associate the presence of a movie with individual interactions or recommendations based on time: we consider a book to be an \textbf{active movie-adapted book} if the interaction or recommendation occurs in a 6-month window beginning 1 month before the movie release to 5 months after (to account for both anticipatory and lingering effects of movie release).
Further detail and justification of this choice is in the next section.

\section{Movie Release and Book Readership}
\label{sec:rq1}
To address \textbf{RQ1}, we conduct a descriptive empirical analysis combining event-study visualizations and a Differences-in-Differences (DiD) framework \cite{angrist2009mostly} to characterize readership dynamics around movie release dates. While not intended to establish causal identification, this analysis provides structured evidence on temporal patterns and heterogeneity in engagement following movie releases.

\paragraph{Event-Study Design.} To investigate how movie releases affect book readership, we discard books without sufficient observations during the pre-release period required for the analysis, yielding a final set of 261 adapted books. For each adapted (treated) book, we construct a matched control group of up to $k=5$ non-adapted books from the Goodreads catalog using nearest-neighbor matching on log-transformed pre-release readership \citep{rubin1973matching,stuart2010matching}, with a caliper of 0.5 log units. To account for heterogeneity across books, we further partition the treated books into 10 deciles based on pre-release popularity, measured as the monthly number of new Goodreads readers normalized by platform size. Finally, each matched control is aligned to the same calendar months as its corresponding treated book to ensure comparable readership trajectories.
To characterize readership dynamics around movie releases, we aggregate Goodreads interactions into monthly counts and align each book by its release date. For every movie-adapted title, we define three consecutive six-month windows relative to the release month: a \emph{pre-release} period $[-7,-1)$, an \emph{event} period $[-1,+5]$, and a \emph{post-release} period $(+5,+11]$, with the event window beginning one month prior to release to account for anticipation effects.
Matching reduces the standardized mean difference (SMD) in pre-release readership from 2.63 to 0.22, improving balance between groups.

\paragraph{Event-Study Analysis.} Figure \ref{fig:event_study} reports the average monthly number of new readers per movie-adapted book compared to their matched controls, stratified by deciles of pre-release book readership; raw interaction counts yield equivalent patterns and are omitted for brevity. Stratifying along the matching variable ensures that treated (i.e., movie-adapted) and control books within each panel are maximally comparable by construction, yielding especially clean counterfactual comparisons. Across all ten popularity deciles, movie-adapted books exhibit a pronounced increase in new readers around the movie release that is not observed among matched controls. This event-related spike is consistently observed across deciles, although post-release trajectories differ, with some exhibiting sustained gains and others a more rapid attenuation. Although the magnitude varies across popularity levels, the pattern is evident even in the lowest deciles, suggesting that the movie release effect on book readership is broad and systematic rather than confined to already-prominent titles.

\paragraph{Movie-Release Effect.} To quantify the observed effect, we estimate a Difference-in-Differences (DiD) model \citep{lechner2011estimation} at the treated-book level: for each of the 261 treated books, we average readership across its up to five matched controls to form a single counterfactual, then compute the double difference between the treated book's event-versus-pre-release change in readership and the corresponding change for its averaged control.
This estimator relies on the parallel-trends assumption --- that treated and control readership would have followed the same trajectory absent the release --- which we assess via the placebo test below.
Averaging this per-book estimate across all treated books and testing it against zero with a one-sample $t$-test (so each book contributes one independent observation, despite being matched to up to five controls) yields an estimated increase of 143,989 new Goodreads readers per treated book ($t=6.39$, $p<0.001$).


To assess whether the estimated association reflects pre-existing temporal trends rather than the movie release itself, we conduct a placebo test by shifting the intervention date twelve months earlier and re-estimating the DiD specification \citep{roth2023s}. The resulting placebo estimate is not statistically different from zero (DiD = ${-48{,}617}$, $p = 0.14$), suggesting limited evidence of differential pre-trends in the outcome prior to treatment.

\begin{figure*}[ht]
    \centering
  \begin{subfigure}{0.32\textwidth}
    \includegraphics[width=1\linewidth]{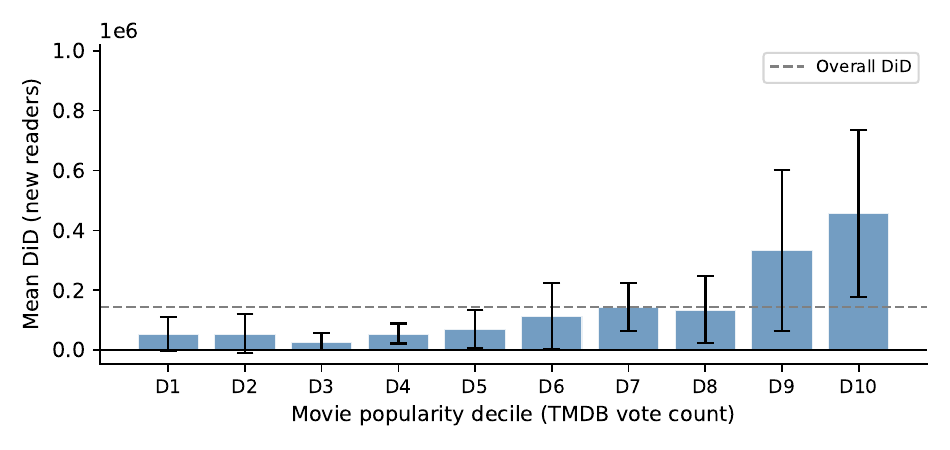}
    \caption{Movie popularity decile}\label{fig:het-movie}
  \end{subfigure}\hfill
  \begin{subfigure}{0.32\textwidth}
    \includegraphics[width=1\linewidth]{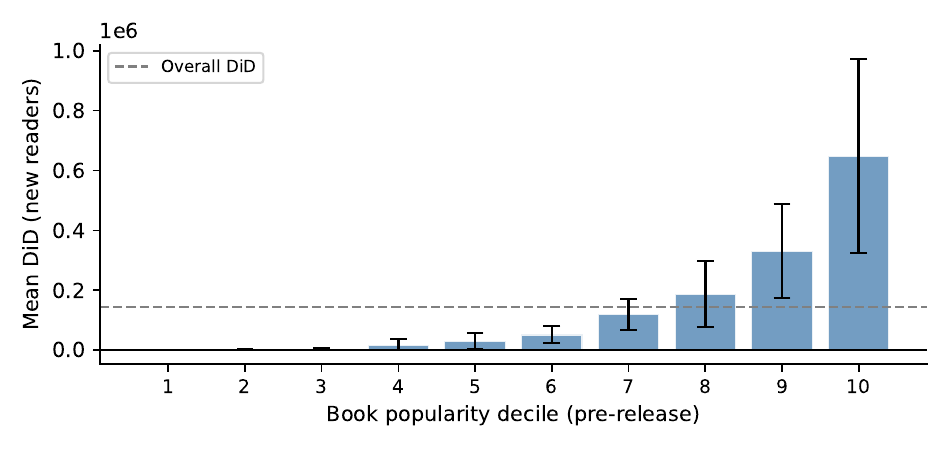}
    \caption{Book decile, absolute DiD}\label{fig:het-book-abs}
  \end{subfigure}\hfill
  \begin{subfigure}{0.32\textwidth}
    \includegraphics[width=1\linewidth]{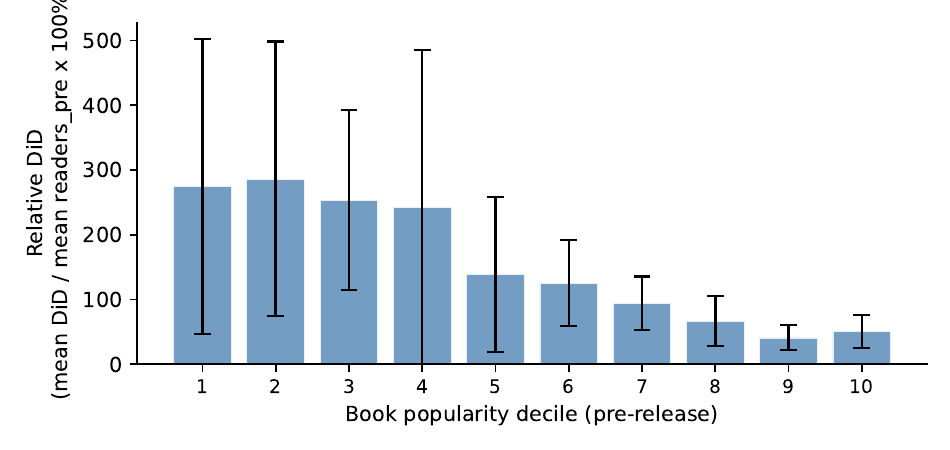}
    \caption{Book decile, relative DiD}\label{fig:het-book-rel}
  \end{subfigure}
  \caption{Heterogeneous DiD estimates by movie and book popularity (95\%
confidence intervals). (\subref{fig:het-movie}) Mean DiD by movie popularity
decile (TMDB votes). (\subref{fig:het-book-abs}) Mean DiD by pre-release book
popularity decile. These panels share a common linear $y$-axis. 
(\subref{fig:het-book-rel}) Relative DiD by pre-release book popularity decile,
expressed as a percentage of mean pre-release readership. The dashed grey line
marks the overall mean DiD.}
  \label{fig:het}
\end{figure*}

\paragraph{Heterogeneous Effects.}
Figure~\ref{fig:het} reports DiD estimates stratified by movie popularity decile
(Fig.~\ref{fig:het-movie}) and pre-release book popularity decile
(Figs.~\ref{fig:het-book-abs} and~\ref{fig:het-book-rel}), together with 95\% confidence intervals. Since each decile contains approximately 25-30 treated books, individual estimates are associated with substantial uncertainty. Consequently, we focus on the overall trends across deciles rather than the magnitude of individual estimates.
Across pre-release book popularity deciles (Figs.~\ref{fig:het-book-abs} and~\ref{fig:het-book-rel}), the estimated effect increases monotonically in absolute terms. In relative terms, however, the pattern reverses: books with limited prior readership exhibit proportional increases approximately five to six times larger than those observed for the highest-popularity decile.
Across movie popularity deciles (Fig.~\ref{fig:het-movie}), the estimated effect is consistently positive and generally increases with movie popularity, although the relationship appears less monotonic than for book popularity. The largest gains are observed for the two highest movie-popularity deciles.

Overall, the observed readership dynamics suggest that movie-adaptation status constitutes a potentially informative signal for recommendation. We next investigate whether this signal is already reflected in the scores assigned by standard recommender models.

\section{Movie Release and Recommendation Scores}

To address \textbf{RQ2}, we study whether recommender models account for movie release signals by analyzing both the ranking of active movie-adapted books in candidate lists and the scores assigned to them relative to comparable non-adapted books.

We use four recommenders: Implicit Matrix Factorization (ImplicitMF) \citep{hu2008collaborative}, Bayesian Personalized Ranking (BPR) \citep{rendle2012bpr}, a time bounded popularity baseline (Pop), and SLIM \citep{ning2011slim}, with hyperparameters tuned on a temporally distinct 2013 validation window.
All models are trained on interactions before January 1, 2017 and evaluated on the following two months.
Pop's scores use a trailing 12-month popularity window (2016-01 to 2017-01).
Each model generates a top-2500 candidate list per user.
We implement these models using LensKit \citep{ekstrand2020lenskit}.



Table~\ref{tab:stats} reports retrieval and ranking statistics for the $33,631$ interactions involving the $24$ active movie-adapted books in the 2017-01 evaluation window.
A notable fraction of these interactions involve books not retrieved into the candidate list, and those that are retrieved tend to be ranked low.
Pop retrieves the highest share of interactions but concentrates them near the top, reflecting its bias toward globally popular books.
Personalized models retrieve a smaller share overall but achieve higher top-100 rates, suggesting they surface a broader range of movie-adapted books.
These results show that active movie-adapted books are frequently missed or buried, but do not tell us whether this reflects lower user affinity or systematic score differences. 

\subsection{Movie Release-Effect Estimation}
\label{sec:causal-method}
We investigate whether recommender systems systematically assign different scores (and therefore rankings) to books with recent or upcoming movie adaptations by framing the problem as an observational causal inference task and estimating the average treatment effect on the treated (ATT) of movie-adaptation status on recommendation scores. Consistent with the recommender inference window, treatment is defined at the book level as the release of a movie adaptation within the $[-1,+5]$-month window relative to the recommendation date. The outcome is the recommendation score assigned to each user-book pair. Since recommendation scores are model-specific and not directly comparable across recommenders, all analyses are conducted independently for each model. Although the analysis is framed within the potential-outcomes framework \cite{imbens2015causal}, we interpret the estimated effects conservatively as covariate-adjusted associations, with a causal interpretation contingent on the assumptions and limitations discussed at the end of this section.

\begin{table}[]
\caption{Interaction statistics for active movie-adapted books.}
\label{tab:stats}
\begin{tabular}{lrrrr}
\toprule
& ImplicitMF & BPR & Pop & SLIM\\
\midrule
Retrieved in top-2500 & 78.7\%  & 76.6\% & 91.1\% & 71.2\% \\
Median rank & 255 & 301 & 256 & 286 \\
Mean rank & 513 & 574 & 436 & 541 \\
In top-10 & 6.4\% & 3.3\% & 15.6\% & 4.7\% \\
In top-100 & 30.8\% & 27.5\% & 26.9\% & 27.6\% \\
\bottomrule
\end{tabular}
\end{table}

\paragraph{Control Pool and Matching} To address observed confounding, we implement a two-stage matching procedure:
coarsened exact matching (CEM) \cite{iacus2008matching, iacus2012causal} for
exact stratification on genre and coarsened bins of book age and pre-release
popularity, followed by a within-stratum $k$-nearest-neighbor refinement
($k=5$) without replacement, using standardized distances on the uncoarsened
covariates to reduce residual imbalance within each coarsened cell
\cite{iacus2009cem, ho2007matching}. To avoid contamination, we exclude
candidate controls whose own movie adaptation was released before the
treatment window (plus a one-month buffer), as they may already carry a
movie-related signal at scoring time; books missing matching covariates are
also discarded. Under this design, all $24$ treated books are matched,
yielding $111$ controls ($4.6$ per treated book on average), with excellent
covariate balance, with SMD reaching $0.018$.

\paragraph{Estimand and Estimator}
We estimate the ATT of movie-adaptation status, weighting each treated book
equally. Let
$T_i\in\{0,1\}$ indicate whether book $i$ is a movie adaptation and
$s_{u,i}(1),s_{u,i}(0)$ the potential recommendation scores for pair $(u,i)$
under the adapted and non-adapted states of book $i$, with observed score
$s_{u,i}=s_{u,i}(T_i)$. Let $\mathcal{S}$ be the set of scored $(u,i)$ pairs.
For each of the $G$ matched groups $g$ (treated book $t(g)$, matched controls
$C(g)$), let $C_u(g)=\{c\in C(g):(u,c)\in\mathcal{S}\}$ be the controls scored
for user $u$ and $U_g$ the users for whom $t(g)$ and at least one control are
scored. Let
\begin{equation}
\tau_g =
\mathbb{E}_u\left[
s_{u,t(g)}(1)-s_{u,t(g)}(0)
\mid u\in U_g
\right]
\end{equation}
be the average treatment effect for matched group $g$ over users for whom the
treated book and at least one control are scored. The target estimand is the
equally weighted average across matched groups,
\(\tau_{\mathrm{ATT}}
=
\frac{1}{G}\sum_{g=1}^{G}\tau_g .\)
For each matched group, we estimate the group-level effect as:
\begin{equation}
\widehat{\tau}_g =
\frac{1}{|U_g|}
\sum_{u\in U_g}
\left(
s_{u,t(g)}
-
\frac{1}{|C_u(g)|}
\sum_{c\in C_u(g)} s_{u,c}
\right),
\end{equation}
and aggregate these estimates as
\(\widehat{\tau}_{\mathrm{ATT}}
=
\frac{1}{G}
\sum_{g=1}^{G}
\widehat{\tau}_g .\)
Under conditional ignorability and overlap, the matched-control average
estimates the counterfactual $s_{u,t(g)}(0)$, and differencing within $u$
removes user-specific score tendencies. Since treatment is assigned at the book level, uncertainty should account for
clustering at the matched-group level \citep{abadie2023should}. We therefore implement
a block bootstrap that resamples the $G$ matched groups with replacement
\citep{cameron2008bootstrap}, recomputing the ATT at each replicate from the resampled
group-level estimates $\widehat{\tau}_g$. This procedure captures
between-book rather than within-group variability. Given the small number of
matched groups ($G\approx24$), inference in this few-cluster setting may have
limited coverage accuracy; we therefore interpret the resulting confidence
intervals as indicative. For cross-model comparability we report a
standardized effect $\widehat{\tau}_{\mathrm{ATT}}/\widehat{\sigma}_{C}$,
where $\widehat{\sigma}_{C}$ is the pooled standard deviation of
matched-control scores.

\paragraph{Results}
\begin{table}[t]
\caption{Per-model book-equal ATT of movie-adaptation status on the relevance
score, in native ($\widehat{\tau}_{\mathrm{ATT}}$) and standardized
($\widehat{\tau}^{\,\mathrm{std}}_{\mathrm{ATT}}$, $\div$ pooled control-score SD)
units, with group block-bootstrap 95\% CIs. $G$: number of matched groups.}
\label{tab:att}
{\centering\small\setlength{\tabcolsep}{5pt}
\begin{tabular}{lccc}
\toprule
Model & $\widehat{\tau}_{\mathrm{ATT}}$ [95\% CI] & $\widehat{\tau}^{\,\mathrm{std}}_{\mathrm{ATT}}$ & $G$ \\
\midrule
ImplicitMF & $0.010$ $[-0.002,\,0.023]$ & $0.047$ & $24$ \\
BPR        & $0.089$ $[-0.039,\,0.213]$ & $0.062$ & $24$ \\
SLIM       & $0.005$ $[-0.002,\,0.012]$ & $0.040$ & $24$ \\
Pop  & $0.040$ $[-0.000,\,0.082]$ & $0.528$ & $15$ \\
\bottomrule
\end{tabular}}
\end{table}

Table~\ref{tab:att} reports the ATT estimates. For the three personalized models, the estimated effects are small and their 95\%
confidence intervals include zero. Thus, we find no evidence that these
recommenders systematically assign higher or lower scores to recently adapted books
than to comparable non-adapted books. The popularity baseline
yields the largest standardized value, but this reflects its low control-score
dispersion (being non-personalized, its scores are near-constant across
users) and its interval likewise includes
zero. Moreover, the estimate is based on fewer matched groups ($G=15$),
which further limits the reliability of its inference. 

\paragraph{Assumptions and Limitations}
The ATT estimates have a causal interpretation under conditional ignorability, overlap, and SUTVA.
Our design relies on the assumption that, conditional on the matched pre-treatment covariates, adaptation status is independent of potential recommendation scores. While models score books independently,
SUTVA is only approximately satisfied because training relies on a shared
interaction matrix that may already capture adaptation-driven engagement,
introducing limited spillovers.  Moreover, movie adaptations are heterogeneous in release, marketing, and popularity, and some upcoming adaptations may not yet be reflected in the model. Matching balances observed
covariates but cannot rule out residual confounding from unobserved factors such
as author prominence or marketing exposure. Finally, the limited number of
treated books reduces statistical power and makes confidence intervals
less reliable. 

\section{Conclusion}
In this paper, we explored how movie releases affect book readership and whether existing recommender systems respond to such exogenous events and demand shifts.
We found that movie releases were associated with measurable readership spikes in adapted books.
We do not see statistical evidence that collaborative filtering models account for (or fail to account for) movie release impact on book consumption, although a modest effect cannot be ruled out.
These findings motivate event-aware strategies: adding adaptation status and time-to-release as temporal item features so models learn a release-proximity effect directly, or applying a post-hoc re-ranking boost during the release window.
Since individual events are rare at inference time, evaluating such strategies likely requires pooling across many historical release windows rather than relying on a single event for power.

\begin{acks}
This work was supported by the National Science Foundation under Grant IIS 24-09199.
\end{acks}

\bibliographystyle{ACM-Reference-Format}
\bibliography{sample-base}

\appendix

\end{document}